\def \SAIT #1 #2 {{\em Mem.\ Soc.\ Astron.\ It.\/} {\bf #1}, #2}
\def \MESS #1 #2 {{\em The Messenger\/} {\bf #1}, #2}
\def \ASTRNACH #1 #2 {{\em Astron. Nach.\/} {\bf #1}, #2}
\def \AAP #1 #2 {{\em Astron. Astrophys.\/} {\bf #1}, #2}
\def \AAL #1 #2 {{\em Astron. Astrophys. Lett.\/} {\bf #1}, L#2}
\def \AAR #1 #2 {{\em Astron. Astrophys. Rev.\/} {\bf #1}, #2}
\def \AAS #1 #2 {{\em Astron. Astrophys. Suppl. Ser.\/} {\bf #1}, #2}
\def \AJ #1 #2 {{\em Astron. J.\/} {\bf #1}, #2}
\def \ANNREV #1 #2 {{\em Ann. Rev. Astron. Astrophys.\/} {\bf #1}, #2}
\def \APJ #1 #2 {{\em Astrophys. J.\/} {\bf #1}, #2}
\def \APJL #1 #2 {{\em Astrophys. J. Lett.\/} {\bf #1}, L#2}
\def \APJS #1 #2 {{\em Astrophys. J. Suppl.\/} {\bf #1}, #2}
\def \APSS #1 #2 {{\em Astrophys. Space Sci.\/} {\bf #1}, #2}
\def \ASR #1 #2 {{\em Adv. Space Res.\/} {\bf #1}, #2}
\def \BAIC #1 #2 {{\em Bull. Astron. Inst. Czechosl.\/} {\bf #1}, #2}
\def \JSQRT #1 #2 {{\em J. Quant. Spectrosc. Radiat. Transfer\/} {\bf #1}, #2}
\def \MN #1 #2 {{\em Mon. Not. R. Astr. Soc.\/} {\bf #1}, #2}
\def \MEM #1 #2 {{\em Mem. R. Astr. Soc.\/} {\bf #1}, #2}
\def \PLR #1 #2 {{\em Phys. Lett. Rev.\/} {\bf #1}, #2}
\def \PASJ #1 #2 {{\em Publ. Astron. Soc. Japan\/} {\bf #1}, #2}
\def \PASP #1 #2 {{\em Publ. Astr. Soc. Pacific\/} {\bf #1}, #2}
\def \NAT #1 #2 {{\em Nature\/} {\bf #1}, #2}
\def  \deg {{$^{\circ}$}}

\documentstyle{memsait}
\input epsf.sty
\begin{opening}
\title{THE ARTIFICIAL SKY LUMINANCE AND THE EMISSION ANGLES OF THE UPWARD LIGHT FLUX} 
\author{ PIERANTONIO CINZANO$^1$, FRANCISCO JAVIER DIAZ CASTRO$^2$}
\institute{$^1$Dipartimento di Astronomia, Universit\`a di Padova,
vicolo dell'Osservatorio 5,\\  I-35122 Padova, Italy\\
email: cinzano@pd.astro.it\\
$^2$Oficina T\'ecnica para la Protecci\'on de la Calidad del Cielo, Instituto de Astrofisica de Canarias, La Laguna, Tenerife, Spain\\
email: fdc@ll.iac.es}
\date{} 
\end{opening}

\begin{document}

\oddpagefooter{}{}{} 
\evenpagefooter{}{}{} 
\ 
\bigskip

\begin{abstract}
The direction of the upward light emission has different polluting effects on the sky depending on the distance of the observation site. We studied with detailed models for light pollution propagation the ratio $\frac{b_{H}}{b_{L}}$, at given distances from a city, between the artificial sky luminance $b_{H}$ produced by its upward light emission between a given threshold angle $\theta_{0}$ and the vertical and the artificial sky luminance $b_{L}$ produced by its upward light emission between the horizontal and the threshold angle $\theta_{0}$. Our results show that as the distance from the city increases the effects of the emission at high angles above the horizontal decrease relative to the effects of emission at lower angles above the horizontal. Outside some kilometers from cities or towns the light emitted between the horizontal and 10\deg ~is as important as the light emitted at all the other angles in producing the artificial sky luminance. Therefore
the protection of a site requires also a careful control of this emission which needs to be reduced to at most 1/10 of the remaining emission. The emission between the horizontal and 10\deg ~is mostly produced by spill light from luminaires, so fully shielded fixtures (e.g. flat glass luminaires or asymmetric spot-lights installed without any tilt) are needed for this purpose.

An adequate protection of the sky with the aim to save citizen's capability to see the heavens near or inside cities and towns requires the limitation of the upward emission at all emission angles.
Nevertheless a specific limitation of light emitted at low angles over the horizontal is useful also in highly urbanized areas where an important fraction of the artificial sky luminance, even inside mean-size cities, is produced by the sum of the contributions of a big number of sources in the surrounding land. The use of fully shielded fixtures can greatly reduce this contribution from outside. 
\end{abstract}

\section{Introduction}

The light pollution produced by the upward light emission from night-time lighting  causes a sky glow which disturbs and, sometime prevents, both astronomical observations and the capability of citizens to see the night sky, a world heritage. 

Even if all the upward light emission causes pollution, the direction of this emission has different effects depending on the distance of the observation site.

We studied with detailed models for light pollution propagation the ratio $\frac{b_{H}}{b_{L}}$, at given distances from a city, between the artificial sky luminance $b_{H}$ produced by its upward light emission between a given threshold angle $\theta_{0}$ and the vertical and the artificial sky luminance $b_{L}$ produced by its upward light emission between the horizontal and the threshold angle $\theta_{0}$. In section \ref{sec2} we summarize main steps in modelling the ratio. In section \ref{sec3} we present our results. The conclusions are in section \ref{concl}.

\section{The modelling technique}
\label{sec2}
The models are based on the modelling technique introduced and developed by Garstang (1986, 1987, 1988, 1989a, 1989b, 1989c, 1991a, 1991b, 1991c, 1992, 1993, 1999) and applied in Italy by Cinzano (1999a, 1999b). 
Main steps are:
\begin{enumerate}
\item For each infinitesimal volume of atmosphere along the line-of-sight, the direct illuminance produced by the city and the illuminance due at light scattered once from molecules and aerosols are computed. The model assumes Rayleigh scattering by molecules and Mie scattering by aerosols.  
\item The total flux that molecules and aerosols in the infinitesimal volume scatter toward the observer are computed from the illuminance
\item The artificial sky luminance of the sky in that direction is obtained with an integration. 
\end{enumerate}
Extinction along light paths is taken in account. 
The same atmospheric model as Garstang (1986, 1991) was assumed, with the density of molecules and aerosols decreasing exponentially with the height and the average scattering function of aerosols as measured by McClatchey et al. (1978).
Computations has been done for standard clear atmosphere (Garstang 1996).
We neglected the curvature of the Earth.

\begin{figure}
\epsfysize=5cm 
\hspace{1.0cm}\epsfbox{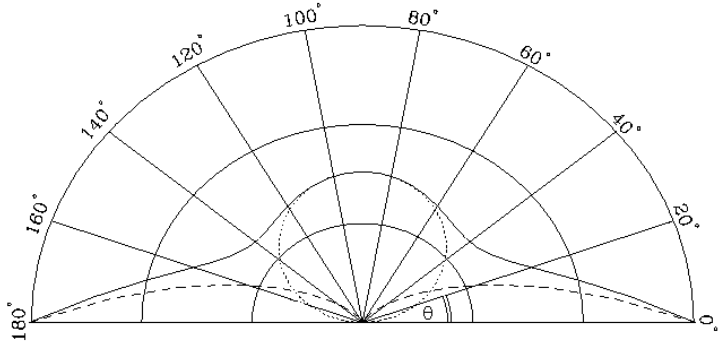} 
\caption[h]{Average city emission function from Garstang (1986).}
\label{fig1}
\end{figure}

The  average emission function of cities  is the sum of the direct emissions from fixtures and the reflected emissions from lighted  surfaces and is not well known. 
In a first set of models we used 
the average city emission function of Garstang (1986) (G=0.15, F=0.15). 
The function is shown in figure \ref{fig1} together with the definition of the emission angle $\theta$.
We used the Garstang function considering it like a parametric expression which is a good approximation of the total city emission function, as stated by many successful comparisons of models to observational data (see cit.). We didn't give to its components any of the meanings gived by that author. We discarded the hypothesis that all surfaces emit only as a vertical Lambertian and that all fixtures emit only as $\theta^{4}$ which was used by Garstang to introduce this expression.
We considered parameters G and F only as shape parameters without any connection with reflected or directly emitted light. In a second set of models we used an isotropic city emission function for comparison purposes.
When the line of sight approaches the city closer than 12
times its radius we use for it a  seven points approximation (Abramowitz and Stegun
1964). 

\section{Results}
\label{sec3}
\begin{figure}
\epsfysize=15cm 
\hspace{1.5cm}\epsfbox{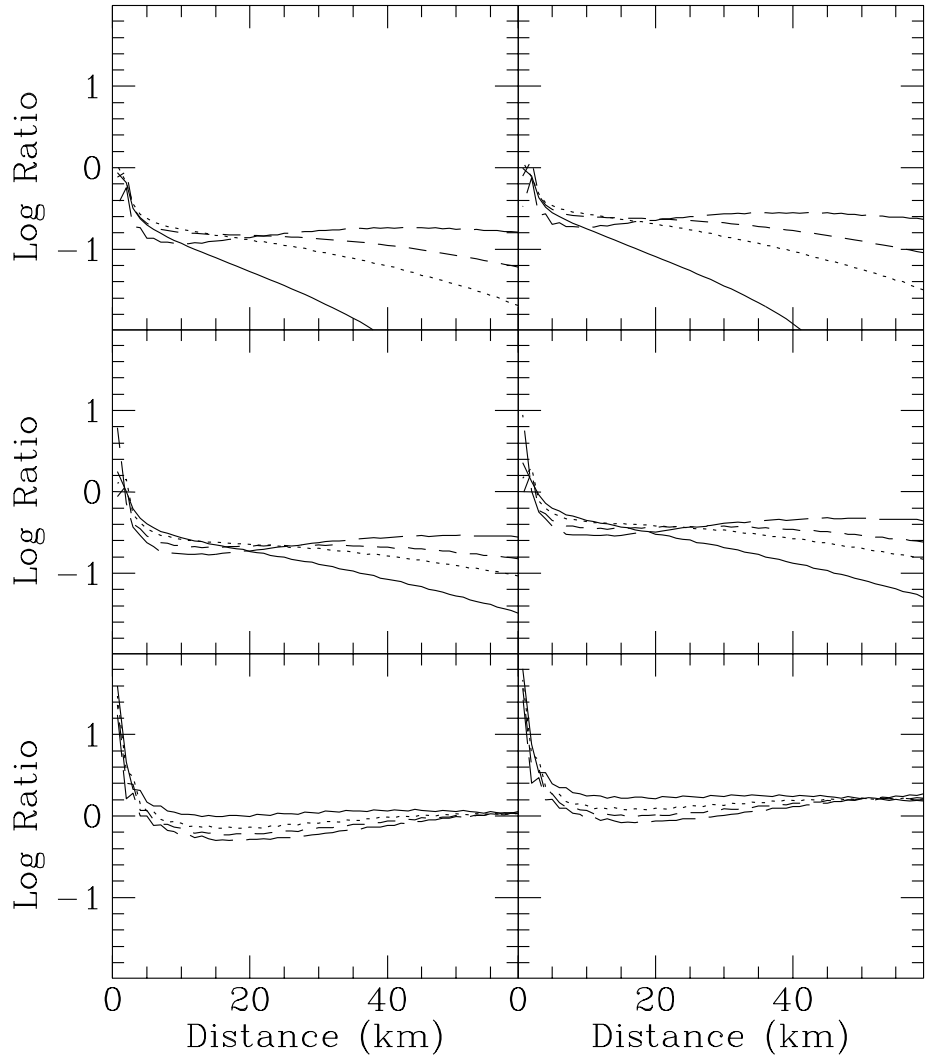} 
\caption[h]{Ratios $\frac{b_{H}}{b_{L}}$ between the artificial sky luminance $b_{H}$ produced by  upward light emission of a city between the threshold angle $\theta_{0}$ and the vertical and the artificial sky luminance $b_{L}$ produced by its upward light emission between the horizontal and the threshold angle $\theta_{0}$. The ratios are computed for threshold angles $\theta_{0}$ of 45\deg (top panel), 30\deg (middle panel) and 10\deg (bottom panel). Left panels show ratios for models with the Garstang city emission function (G=0.15, F=0.15), right panels show ratios for models with isotropic city emission function. Curves refer to zenith distances of 0\deg (solid curves), 30\deg (dotted curves), 45\deg (dashed curves), 60\deg (long dashed curves) towards the city.}
\label{curves}
\end{figure}

Figure \ref{curves} shows the ratio $\frac{b_{H}}{b_{L}}$, at given distances from a city, between the artificial sky luminance $b_{H}$ produced by its upward light emission between the threshold angle $\theta_{0}$ and the vertical and the artificial sky luminance $b_{L}$ produced by its upward light emission between the horizontal and the threshold angle $\theta_{0}$. The ratios are computed for threshold angles $\theta_{0}$ of 45\deg (top panel), 30\deg (middle panel) and 10\deg (bottom panel). Left panels show ratios for models with the Garstang city emission function (G=0.15, F=0.15), right panels show ratios for models with isotropic city emission function. Curve refers to zenith distances of 0\deg (solid curves), 30\deg (dotted curves), 45\deg (dashed curves), 60\deg (long dashed curves) toward the city.

The figure shows that only inside a city are the sky luminance produced by the emission over 45\deg and that produced by the emission between the horizontal and 45\deg of the same importance. Moving the observation site outward, their ratio decrease quickly. At a distance of $\sim$40 km the contribution near the zenith from light emitted over 45\deg by the city is about 1/100 of that from the light emitted between the horizontal and 45\deg. For higher zenith distances the ratio decreases slower but however it is small.

The sky luminance produced by the emission over 30\deg ~exceeds that produced by the emission between the horizontal and 30\deg ~inside the city boundaries but it is lower again outside. The ratios decrease with the distance from the city slower than in the previous case.

The sky luminance produced by the emission over 10\deg ~is much higher than that produced by the emission between the horizontal and 10\deg ~up to some km from the city. In models with Garstang emission function, the ratio $\frac{b_{H}}{b_{L}}$ near the zenith decreases at about 1 after $\sim$5 km and remains quite constant. In models with isotropic emission function, it decreases to about 2.5 after $\sim$10 km and also remains quite constant. For higher zenith distances the ratios can be slightly lower.

\section{Conclusions}
\label{concl}
Our results show that: 
\begin{enumerate}
\item An adequate protection of the sky with the aim to save citizen's capability to see the heavens near or inside cities and towns requires the limitation of the upward emission at all emission angles.
\item The protection of a site outside some kilometers from cities or towns requires not only a general control of upward flux (both coming from direct emission by fixtures and from reflexion by lightened surfaces) but also a careful control of the light emitted between the horizontal and 10\deg-20\deg. The light emitted in the range 0\deg-10\deg ~is as important as the light emitted at all the other angles in producing the artificial sky luminance there. 
\item Lightened horizontal surfaces have a mixed behaviour between a Lambertian diffusor (emitting very little light at small angles) and a reflector of the light emitted by luminaires (which maximum emission is at 60-70\deg ~from their axis giving a maximum reflection at 20-30\deg). So, the emission between the horizontal and 10\deg ~is mainly produced by direct emission from fixtures. This unnecessary contribution to sky luminance  needs to be reduced to a negligible fraction (at most 1/10 of the remaining emission). So fully shielded fixtures (e.g. flat glass luminaires or asymmetric spot-lights installed without any tilt) are required even at great distances from the site.
\item A specific limitation of light emitted at low angles is useful also in high urbanized areas where the artificial sky luminance is produced by the sum of the contribution of a big number of  sources. From results of modelling sky luminance of the city of Padova (Cinzano 1999) can be inferred that the sky luminance inside a city of less than about 70000 inhabitants in Veneto Plain could be produced more by the sources in the surrounding land than by the city itself. The use of fully shielded fixtures can limit definitely this contribution from outside when the protection of citizen's sky vision capability is wanted. 
\end{enumerate}

\acknowledgements
We are indebted to Roy Garstang of JILA-University of Colorado for his friendly kindness in reading and refereeing this paper, for his helpful suggestions and for interesting discussions.


\end{document}